\documentclass[aps,twocolumn,prl,floatfix]{revtex4-1}

\usepackage[utf8]{inputenc}
\usepackage{amsmath}
\usepackage{amssymb}
\usepackage{amsfonts}
\usepackage{graphicx}
\usepackage{xcolor}
\usepackage{ifthen}

\renewcommand{\Re}{\operatorname{Re}}

\newcommand{\ohm}{\Omega}

\newcommand{\pq}[3][]
{
	{#2}
	\ifthenelse{\equal{#2}{} \or \equal{#1}{}}{}{\cdot}
	\ifthenelse{\equal{#1}{}}{}{10^{#1}}
	\ifthenelse{\( \equal{#2}{} \and \equal{#1}{} \) \or \equal{#3}{}}{}{\, }
	\mathrm{#3}
}

\newcommand{\on}[1]{\left(#1\right)}
\newcommand{\Jf}{\nu}
\newcommand{\Rq}{R_{\rm Q}}

\newcommand{\Ic}{I_{\rm C}}
\newcommand{\fsignal}{f_{\rm s}}
\newcommand{\fidler}{f_{\rm i}}

\begin{document}

\title{Quantum limited amplification from inelastic Cooper pair tunneling}

\author{S. \surname{Jebari}$^{1,2}$}
\author{F. \surname{Blanchet}$^{1,2}$}
\author{A. \surname{Grimm}$^{1,2}$}
\author{D. \surname{Hazra}$^{1,2}$}
\author{R. \surname{Albert}$^{1,2}$}
\author{P. \surname{Joyez}$^3$}
\author{D. \surname{Vion}$^3$}
\author{D. \surname{Est\`eve}$^3$}
\author{F. \surname{Portier}$^3$}              
\author{M. \surname{Hofheinz}$^{1,2}$}
\email{email: max.hofheinz@cea.fr}
\affiliation{$^1$ Univ. Grenoble Alpes, INAC-PHELIQS, F-38000 Grenoble, France}
\affiliation{$^2$ CEA, INAC-PHELIQS, F-38000 Grenoble, France}
\affiliation{$^3$SPEC, CEA, CNRS, Université Paris-Saclay
CEA-Saclay 91191 Gif-sur-Yvette, France}

\date{\today}
\maketitle

{\bf Nature sets fundamental limits regarding how accurate the amplification of analog signals may be. For instance, a linear amplifier unavoidably adds some noise which amounts to half a photon at best. While for most applications much higher noise levels are acceptable, the readout of microwave quantum systems, such as spin or superconducting qubits, requires noise as close as possible to this ultimate limit. To date, it is approached only by parametric amplifiers exploiting non-linearities in superconducting circuits and driven by a strong microwave pump tone. However, this microwave drive makes them much more difficult to implement and operate than conventional DC powered amplifiers, which so far suffer from much higher noise.  Here we present the first experimental proof that a simple DC-powered setup allows for amplification close to the quantum limit. Our amplification scheme is based on the stimulated microwave photon emission accompanying inelastic Cooper pair tunneling through a DC-biased Josephson junction, with the key to low noise lying in a well defined auxiliary idler mode, in analogy to parametric amplifiers.}
\begin{figure}
\includegraphics[width=0.99\columnwidth]{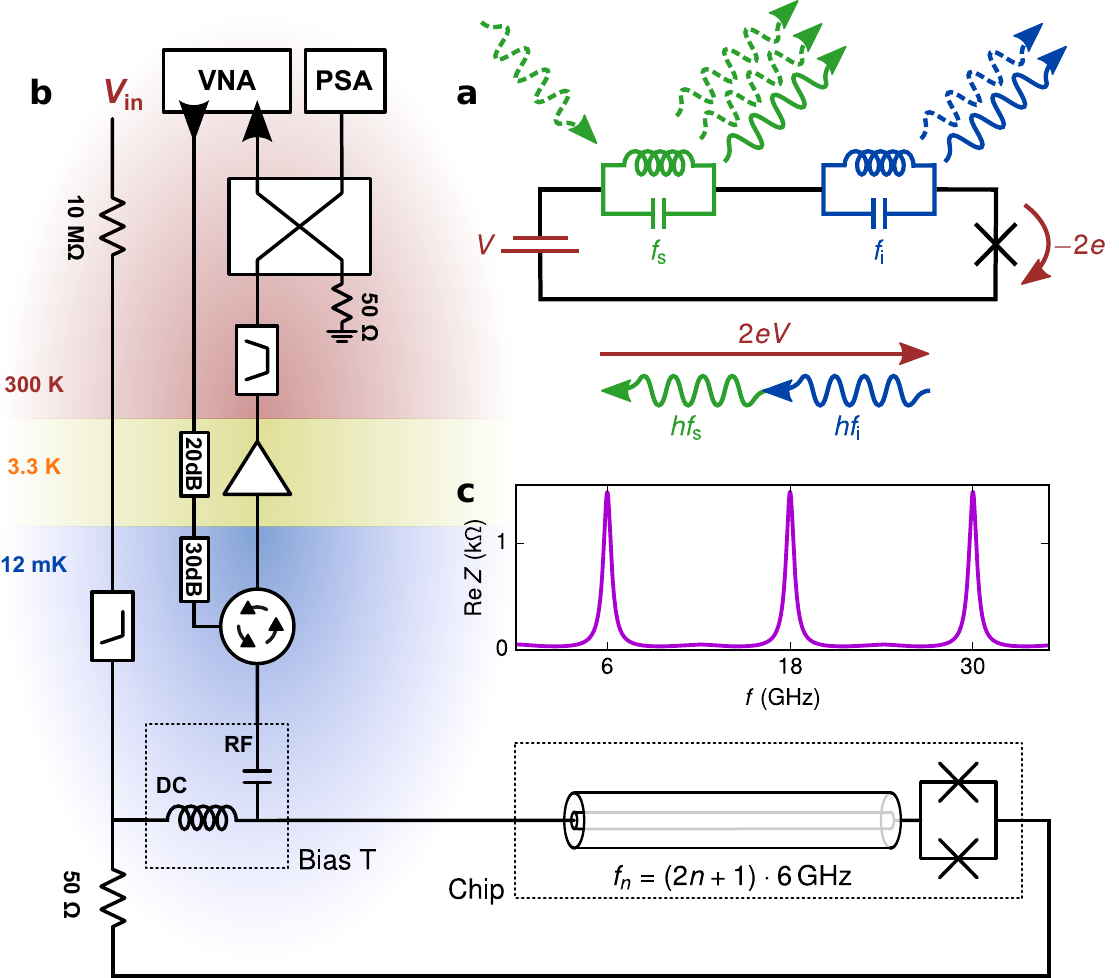}
\caption{\textbf{a, Working principle of an Inelastic Cooper pair Tunneling Amplifier (ICTA):} A Josephson junction ($\times$) in series with two resonators at frequencies $\fsignal$ and $\fidler$ is biased at DC voltage $V$. The energy $2eV$ of a Cooper pair tunneling across the junction can be converted into two photons, one in each resonator (solid wiggly arrows). When a microwave signal is applied at $f \approx \fsignal$, the process accelerates due to stimulated emission. The stimulated response is in phase with the incoming signal (dashed wiggly arrows), giving rise to gain. Photons emitted without incoming signal constitute the unavoidable quantum noise of the amplifier.  \textbf{b, Simplified setup:} The sample consists of an aluminum SQUID, acting as tunable Josephson junction, coupled to a quarter-wave transformer with resonance frequencies $f_n = (2n+1) \cdot \pq{6}{GHz}$ and cooled to $\pq{12}{mK}$ in a dilution refrigerator.  DC voltage is applied through a $\pq{50}{\Omega}$ bias circuit and resonance-free bias T. Microwave reflection and noise are routed via cold circulators (one shown) and measured using a commercial vector network analyzer (VNA) and a custom power spectrum analyzer (PSA). \textbf{c, Impedance $Z(f)$ in series with the SQUID:} It describes the entire linear circuit, including the measurement setup and has maxima at the frequencies $f_n$. Each of these maxima can play the role of the signal or idler mode in the ICTA scheme depicted in a.}
\label{fig:setup}
\end{figure}

\begin{figure*}
\begin{minipage}[c]{0.67\textwidth}
\includegraphics[width=\textwidth]{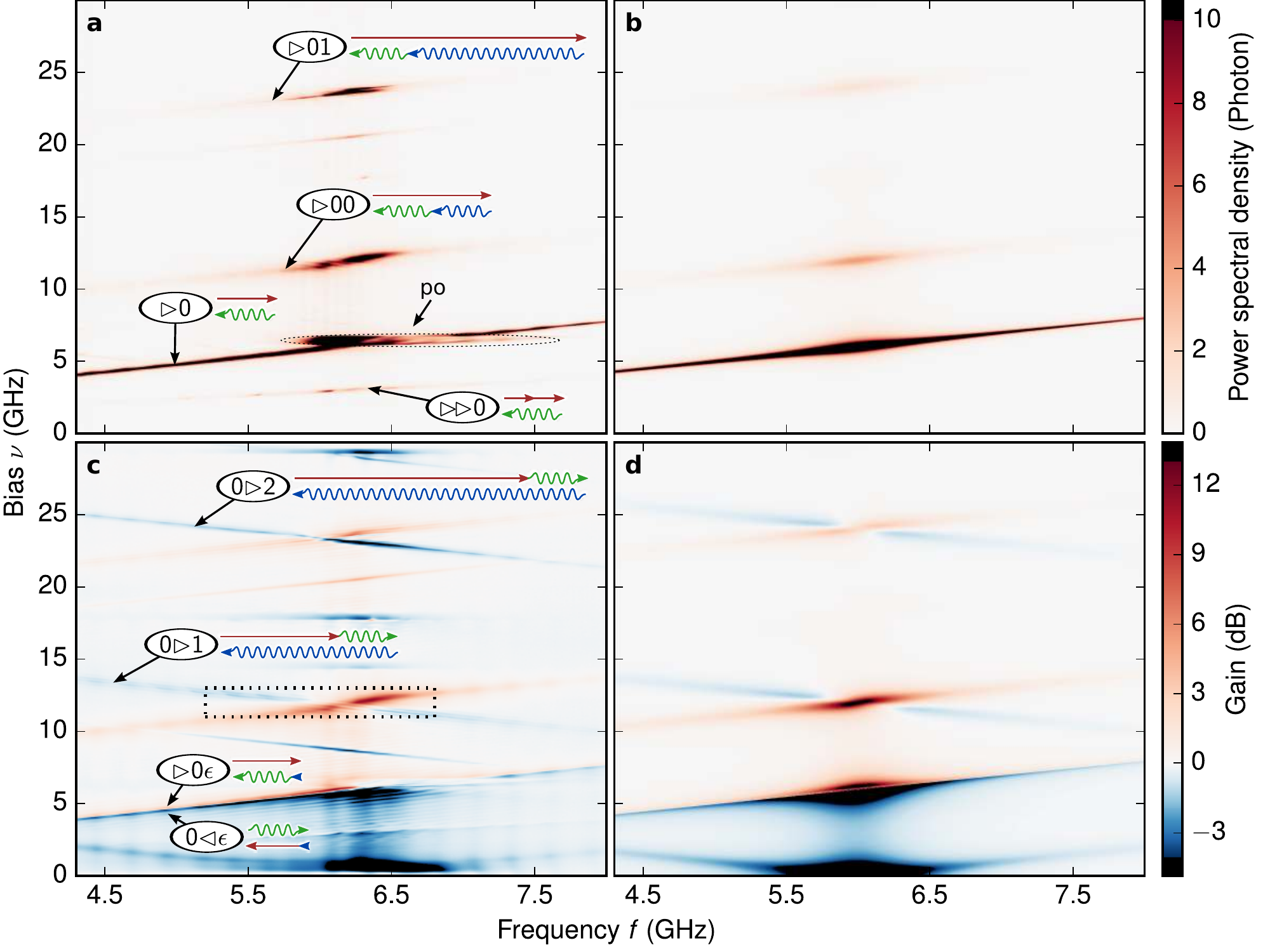}
\end{minipage}\hfill
\begin{minipage}[c]{0.33\textwidth}
  \caption{{\bf a, b, Noise power spectral density (PSD)} measured (a) as a function of Josephson frequency $\Jf$ and frequency $f$ and calculated (b) from $P(E)$ theory (see Supplementary Information) with the designed resonator impedance (see Fig.~\ref{fig:setup}c), $\Ic = \pq{20.2}{nA}$ and an effective temperature of $\pq{54.7}{mK}$. \textbf{c, d, Gain of the ICTA} measured (c) as function of $\Jf$ and $f$ at input signal power $\approx \pq{-117}{dBm}$ and calculated (d) for the same parameters as in b. Red areas correspond to down-conversion processes with gain (amplification) and blue lines to frequency conversion processes observed as loss (see text). The observed inelastic Cooper pair tunneling processes are labelled as follows: ${\rhd}/{\lhd}$ stand for tunneling of a Cooper pair along/against the bias. Symbols before ${\rhd}/{\lhd}$ indicate annihilation of photons, symbols after ${\rhd}/{\lhd}$ indicate creation. Integers $n$ stand for a photon in the mode $f_n$, $\epsilon$ for photons at low frequency. In the energy diagrams next to the labels, red arrows indicate the Cooper pair energy $2eV$, green wiggly arrows the observed photon at frequency $f$ and blue wiggly arrows additional (idler) photons involved.}
\label{fig:psd_gain}
\end{minipage}
\end{figure*}

The quantum limit on the noise of a linear amplifier can be derived from first principles \cite{Caves1982}. This derivation shows that in order to be amplified irrespectively of its phase, the signal necessarily has to be coupled to at least one complementary mode, called idler, and that the photon noise of this mode is added to the signal. In the ideal case, where the idler mode is in its quantum ground state and the gain is large, the added input noise is half a photon. This limit is reached by Josephson parametric amplifiers (JPAs) \cite{Yurke1988,Castellanos-Beltran2007,Bergeal2010,Roch2012,Mutus2014,Roy2015,Macklin2015}, where the nonlinear inductance of Josephson junctions is used to couple a microwave pump tone to the signal mode. They have a perfectly well defined idler mode at the frequency $\fidler$, the difference of (a multiple of) the pump frequency and the signal frequency. This frequency can be matched to a dedicated mode in the circuit which can then be put in its ground state by strongly coupling it to a dedicated cold dissipator with temperature $T \ll h \fidler/k_B$. 

In DC-powered amplifiers, on the other hand, this idler mode is usually not well identified and difficult to engineer. For example, in high electron mobility transistor (HEMT) amplifiers it corresponds to electronic degrees of freedom inside the transistor which are kept out of equilibrium by the DC bias. In DC-powered superconductor-based amplifiers, such as the the superconducting low-inductance galvanometer (SLUG) \cite{Hover2012} or single junction amplifier (SJA) \cite{Lahteenmaki2012,Lahteenmaki2014} the idler can be seen as one of the modes of the dissipative shunt of the junction which also dissipates most of the DC power, so that it gets hot and adds thermal noise.

We implement here a new amplification scheme, which we call Inelastic Cooper pair Tunneling Amplifier (ICTA), based on a Josephson junction biased at DC voltage $V$ below the superconducting gap. A Cooper pair can tunnel through the junction by dissipating its energy $2eV$ in the form of photons \cite{Holst1994, Hofheinz2011}. We focus on processes where this energy is distributed among two photons \cite{Leppakangas2013,Westig2017} (see Fig.~\ref{fig:setup}a). These processes are reminiscent of the parametric down-conversion processes in parametric amplifiers, with the energy $2eV$, playing the role of a pump photon. 

Our device (see Fig.~\ref{fig:setup}b) is the same as in ref.~\cite{Hofheinz2011}. A superconducting quantum interference device (SQUID) acts as a flux-tunable Josephson junction with an estimated maximum critical current $\Ic \sim \pq{20}{nA}$ (see Supplementary Information). It is connected to a $\pq{50}{\Omega}$ transmission line via a quarter-wave transformer yielding resonance peaks in the impedance seen by the junction at $f_n \approx (2n+1) \cdot \pq{6}{GHz}$ with widths of approximately $\pq{500}{MHz}$. We cool the sample down to $\pq{12}{mK}$ and measure its microwave noise emission as well as its microwave reflection (see Fig.~\ref{fig:setup}b).

\begin{figure*}
\begin{minipage}[c]{0.33\textwidth}
  \caption{\textbf{Gain and noise performance.} \textbf{a,} Gain $G$ as in Fig.~\ref{fig:psd_gain}c, but centered around $\nu = \pq{12}{GHz}$. \textbf{b,} Input-referred noise of the ICTA, i.e.\ the measured output noise divided by the gain, taking into account zero-point fluctuations of the input mode. {\bf c,} Cut of data in a at $\Jf = \pq{12.16}{GHz}$ (red) and $ \pq{12.44}{GHz}$ (blue). Dashed lines correspond to calculations based on $P(E)$ theory (see Supplementary Information) for nominal sample parameters, $\Ic = \pq{20.2}{nA}$ and an effective temperature $\pq{54.7}{mK}$. {\bf d,} Cut of data in b at $\Jf = \pq{12.16}{GHz}$ (red) and $ \pq{12.44}{GHz}$ (blue). The dashed lines correspond to the quantum limit $\frac{1}{2}(1-G^{-1})$ of added noise for the measured gain $G$ shown in c.}
\label{fig:gain_noise}
\end{minipage}\hfill
\begin{minipage}[c]{0.66\textwidth}
\includegraphics[width=\columnwidth]{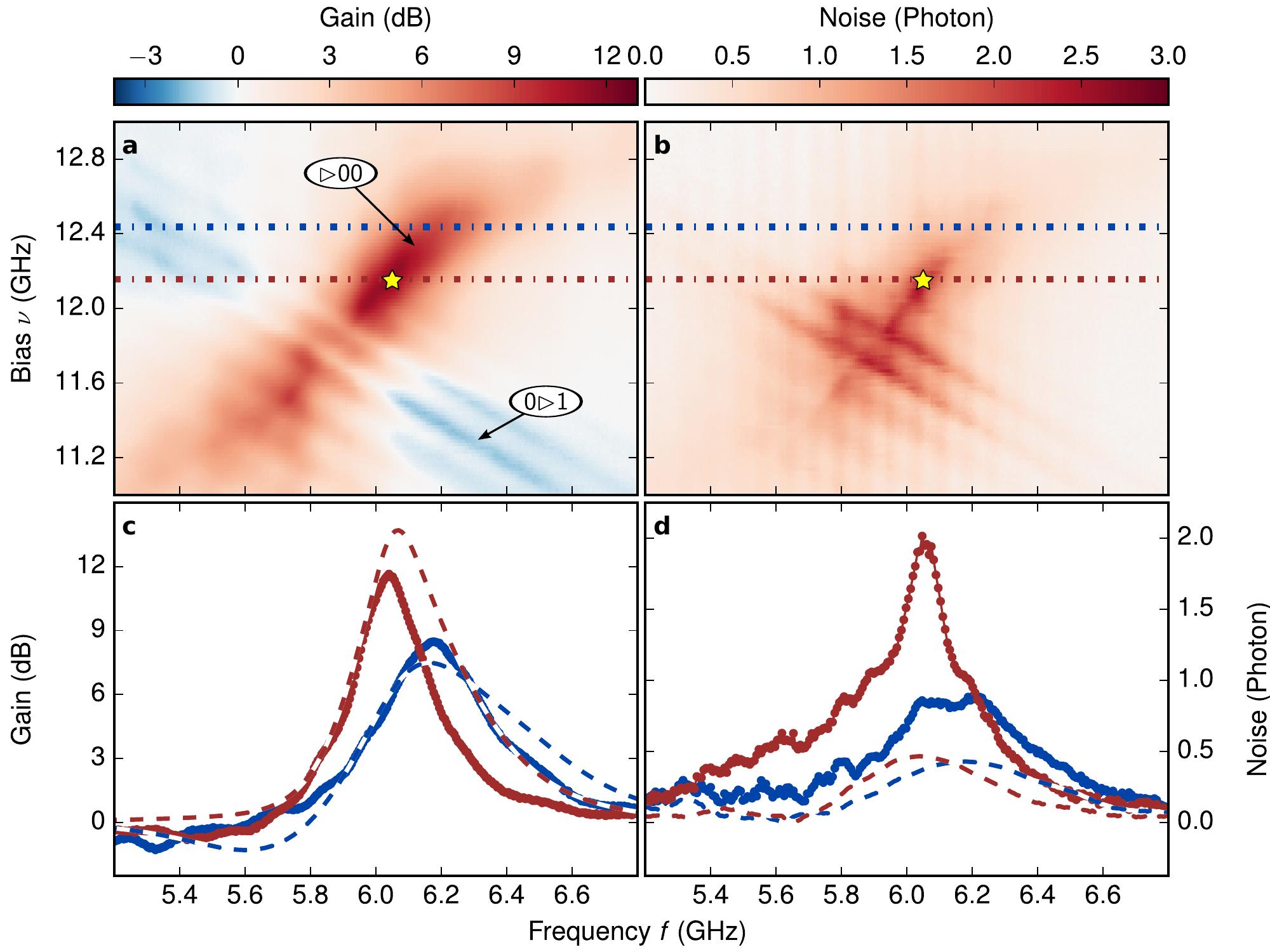}
\end{minipage}
\end{figure*}

In Fig.~\ref{fig:psd_gain}a we show, when no signal is fed in, the noise spectrum emitted by the sample as function of frequency $f$ and applied bias voltage, expressed in terms of Josephson frequency $\nu=2eV/h$. The strongest noise signature appears along the line $\nu = f$ (label $\rhd0$) where the energy $2eV$ of a tunneling Cooper pair is transformed into one photon at $f$. This inelastic Cooper pair process is triggered by zero-point fluctuations of the phase at $f$ which are proportional to $\Re Z(f)/f$ \cite{Ingold1992, Holst1994, Basset2010}. Therefore it is strongest around the impedance maximum $f_0 = \pq{6}{GHz}$.

Additional signatures appear at $\nu = f + f_{0,1}$ (labels $\rhd00$ and $\rhd01$). They correspond to processes where the energy of a tunneling Cooper pair is distributed among two photons. Here the process depends on the zero-point fluctuations at the two frequencies involved. One of the two photons is observed at frequency $f$. Therefore the intensity is again highest at $f\approx f_0$. The other photon involved is likely to be emitted at a frequency where the impedance is high, i.e.\ into any of the modes $f_n$. Therefore the two-photon signatures are shifted by $f_n$ with respect to the one-photon process. 

These two-photon processes are the key ingredient to our amplification process. When triggered by zero-point fluctuations, as discussed so far, they represent the output noise of our amplifier. The same processes can also be triggered by an incoming microwave signal, leading to stimulated emission, in phase with the incoming signal \cite{Safi2011, Roussel2016}. It corresponds to phase-preserving amplification quantified by the gain $G$, the ratio of reflected power over applied power.

In Fig.~\ref{fig:psd_gain}c we show the gain $G$ as a function of signal frequency $f$ and Josephson frequency $\Jf$, when a signal tone of $\approx \pq{-117}{dBm}$ is fed to the device. We indeed observe strong gain, up to $\pq{10}{dB}$, in the areas where we have observed strong two-photon processes in Fig.~\ref{fig:psd_gain}a, meaning that the device indeed provides amplification, as expected.

Figure~\ref{fig:psd_gain}c also shows lines of opposite slope (labelled $0{\rhd}1$ and $0{\rhd}2$) where the device absorbs photons ($G<1$) at the signal frequency even though we expect our device to be essentially dissipationless. These lines can be attributed to another two-photon process where the energy of a tunneling Cooper pair is used to convert an incoming photon at frequency $f$ into a photon at a different frequency $f + \Jf$. The homodyne vector network analyzer (VNA) measurement, however, only detects photons at $f$ and, therefore, the frequency conversion is observed as loss. These two-photon frequency-conversion processes require an incoming photon at frequency $f$ to be present. The absence of these lines in the photon noise in Fig.~\ref{fig:psd_gain}a thus shows that the electromagnetic environment of the device is sufficiently cold to not send any thermal photons at $f$ or $\nu-f$ onto the device.

Along the line $\Jf = f$ both gain (at $\Jf \gtrapprox f$, label ${\rhd}0\epsilon$) and loss (at $\Jf \lessapprox f$, label $0{\lhd}\epsilon$) are visible. They can be explained in the same way as the other gain and loss signatures, but they involve idler photons at very low frequency (see Supplementary Information).

In Fig.~\ref{fig:psd_gain}b and d we compare our measurement with theoretical predictions for noise and gain based on $P(E)$ theory \cite{Ingold1992} (see Supplementary Information). We find a qualitative agreement, correctly describing where we observe gain and loss.

The fact that the gain arises from a down-conversion process triggered by zero-point fluctuations suggests that the ICTA should, in principle, be able to operate at the quantum limit, but just how close can it get to this limit in practice?  To answer this question we focus in Fig.~\ref{fig:gain_noise} on the two-photon process around $\Jf = 2 f_0$. In Fig.~\ref{fig:gain_noise}a we plot the gain $G$ as before. In Fig.~\ref{fig:gain_noise}b we calculate the input-referred noise added by the amplifier by dividing the measured output noise by the measured gain, taking into account zero-point fluctuations of the incoming line (see Supplementary Information). Fig.~\ref{fig:gain_noise}c and d show cuts at $\Jf = \pq{12.44}{GHz}$ and $\pq{12.16}{GHz}$. We observe, respectively, a maximum gain of approximately $\pq{8.5}{dB}$ and $\pq{11.7}{dB}$ over a bandwidth of $\pq{300}{MHz}$ and $\pq{170}{MHz}$. This gain is limited by the critical current of our Josephson junction. We expect that the gain would diverge at less than 2 times higher critical current and then enter a parametric oscillation regime \cite{Chen2014,Cassidy2017} (see Supplementary Information). The input-referred noise at these two bias points is approximately $\pq{0.9}{Photon}$ and $\pq{1.9}{Photon}$. At $\Jf = \pq{12.44}{GHz}$ the input noise indeed corresponds to less than twice the quantum limit (dashed lines in Fig.~\ref{fig:gain_noise}d), a value lower than any existing DC powered amplifier, but higher than the best Josephson parametric amplifiers.

\begin{figure*}
  \begin{minipage}[c]{0.33\textwidth}
    \caption{\textbf{Response at high power}. \textbf{a,} Gain as function of input power for different Josephson energies at $\Jf = \pq{12.15}{GHz}$ and $f=\pq{6.05}{GHz}$ (marked by a star in Fig.~\ref{fig:gain_noise}). The dotted line represents the input $\pq{1}{dB}$ compression points. The dashed line represents Eq.~(\ref{eq:nmax}). \textbf{b,} Gain measured at maximum Josephson energy as function of Josephson frequency and signal frequency for an input power of $P_{\rm in} =\pq{-82}{dBm}$. Overall, gain is strongly compressed and new features appear at $\nu = m f + l f_n$ where $m, l$ integers, corresponding to nonlinear processes involving multiple signal and idler photons.}
\label{fig:high_power}
  \end{minipage}\hfill
  \begin{minipage}[c]{0.66\textwidth}
      \includegraphics[width=\columnwidth]{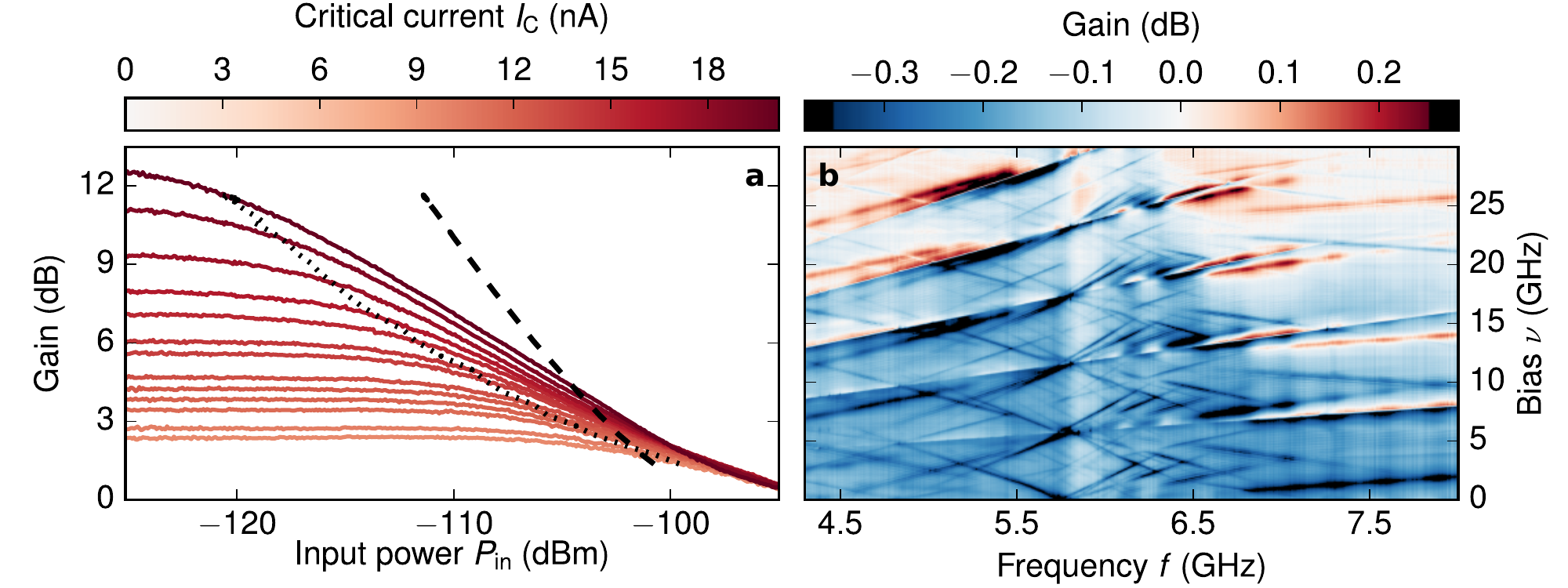}
  \end{minipage}
\end{figure*}

The most straightforward explanations for this excess noise would be losses in the linear circuit and thermal photons in the signal and idler mode, but they can be ruled out: Losses in the circuit are weak and calibrated out (see Supplementary Information) and thermal photons in signal and idler would cause additional signatures in Fig.~\ref{fig:psd_gain}a, as explained above. We instead attribute the excess photon noise to low frequency voltage noise, i.e.\ phase noise of the pump frequency $\Jf$.  If the width of the fluctuations $\Delta \Jf$ is large enough to bring the amplifier out of its optimal working condition, it will modulate the gain in phase and amplitude. In the phase-sensitive VNA measurement (IF bandwidth $\pq{1}{kHz}$) both effects reduce the average gain. The phase insensitive PSD on the other hand is only affected by amplitude fluctuations, so that the input noise, i.e.\ the ratio of photon noise over gain, is degraded.  In our setup we achieve bias fluctuations $\Delta \Jf = \pq{120}{MHz}$ (see Supplementary Information), only slightly lower than the bandwidth of the amplifier at $\Jf = \pq{12.16}{GHz}$. At $\Jf = \pq{12.44}{GHz}$  where the bandwidth is approximately $2\Delta\Jf$ we indeed observe much lower excess noise, in agreement with our explanation. Further confirmation comes from the amplification process labeled ${\rhd} 01$ where we observe lower gain and higher noise than expected. This process involves the mode at $f_1 \approx \pq{18}{GHz}$ which is decomposed in several sharp resonances, due to standing waves in our setup at this frequency (seen as narrow parallel lines in the processes ${\rhd} 01$ and $0{\rhd}1$), making voltage fluctuations particularly harmful.
When the ICTA bandwidth can be made much larger than $\Delta \Jf$ (by increasing the bandwidth and/or reducing $\Delta \Jf$ \cite{Westig2017}), this noise source should become negligible and the ICTA approach quantum limited noise. It should then perform comparably with a JPA where this noise source is essentially absent, because the pump is provided by an external microwave source with negligible phase noise. 

We now address the question what maximum power an ICTA can handle. In Fig.~\ref{fig:high_power}a we show gain compression as a function of input power. At the optimal working point we achieve maximum gain up to an input power of $\pq{-120}{dBm}$ (or $\pq[8]{2.5}{photons/s}$) where the gain drops by $\pq{1}{dB}$. One reason for this gain compression can be understood from Fig.~\ref{fig:high_power}b where we plot the microwave response as a function of voltage and frequency at high power. In addition to the lines at slopes $\pm 1$ involving one photon at $f$, new lines appear at other integer slopes and globally the gain is much lower. These lines correspond to processes involving more than one signal photon, i.e.\ are nonlinear in the signal amplitude. These nonlinearities are avoided when the voltage amplitude is kept below $hf/2e$  \cite{Tien1963,Safi2010}. In addition, the current amplitude is limited by the junction critical current $\Ic$. Together these limits set the maximum input power
\begin{equation}
  P_{\rm in, max} = \frac{hf}{4e} \frac{\Ic}{G-1}
  \label{eq:nmax}
\end{equation}

for the amplifier to stay linear. This limit is close to the observed compression point at low gain (see Fig.~\ref{fig:high_power}a), but unknown effects compress high gain at somewhat lower power. Note that the dynamic range can be improved by embedding the junction in a circuit of relatively low impedance, so that the voltage amplitude at the junction is lower and a higher critical current is needed for the same gain (see Supplementary Information).

In conclusion our results experimentally show that inelastic Cooper pair tunneling can lead to near quantum limited amplification despite imperfections of the bias voltage. We have argued that by designing an appropriate linear matching circuit, characterized by its impedance $Z(f)$ as seen by the Josephson junction, the amplification scheme can be optimized for lower noise, as well as higher bandwidth and dynamic range. Such an amplifier, powered by simple DC voltages could then make measuring microwave signals at the single photon level much easier and allow deploying many amplifiers on a chip. It could, therefore, be an important ingredient for qubit readout in large-scale quantum processors.

\section{Supplementary Information}

\textbf{Gain, loss and parametric oscillation along the one-photon process}
Along the line $\Jf = f$ both gain (at $\Jf \gtrapprox f$, label ${\rhd}0\epsilon$) and loss (at $\Jf \lessapprox f$, label $0{\lhd}\epsilon$) are visible in Fig.~\ref{fig:psd_gain}. They can be explained in the same way as the other gain and loss signatures, but they involve idler photons at very low frequency. Such processes are strong even though the impedance does not have a maximum at low frequency because the probability of photon emission scales with $\Re Z(f)/f$, so that processes involving low frequency photons are enhanced. The loss signature has here the same slope as the gain signature because the Cooper pair tunnels \emph{against} the bias voltage.

This interpretation of the one-photon process as very asymmetric multi-photon process also explains the finite width of the noise signature ${\rhd}0$ in Fig.~\ref{fig:psd_gain}: The low frequency thermal photons can add or remove energy, so that energy $hf$ of the observed photon can be slightly different from the Cooper pair energy $2eV$. The voltage itself is then considered noiseless. In the paper we have, equivalently, included these low frequency thermal fluctuations in the bias voltage which then becomes noisy.

Around $\nu = \pq{6}{GHz}$, $f = \pq{6}{GHz}$, the noise signature in Fig.~\ref{fig:psd_gain}a and the gain signature in Fig.~\ref{fig:psd_gain}b fan out (indicated by label ``po''). We attribute this effect to a parametric oscillation of the amplification process described above\cite{Cassidy2017}. Because of the $\Re Z(f)/f$ scaling, the threshold for parametric oscillation is reached here much earlier than for the desired process involving two photons at $\approx \pq{6}{GHz}$. A much finer scan of this signature shows structure with a spacing of approximately $\pq{35}{MHz}$, indicating the frequency of parametric oscillation. This frequency corresponds to the cross-over frequency of our bias-T where likely a low-Q resonance occurs because the RF branch (see Fig.~\ref{fig:setup}) is not well $\pq{50}{\Omega}$ matched at this frequency.

\textbf{Numerical calculations:} The emitted noise and gain can be qualitatively explained within the $P(E)$ theory of inelastic charge tunneling{\cite{Ingold1992,Holst1994}} and very general relations relating charge tunneling rates and finite frequency noise \cite{Safi2010,Basset2010,Hofheinz2011,Parlavecchio2015,Roussel2016}. The photon emission rate density $\gamma$ at frequency $f$ is proportional to the tunneling rates at shifted voltages:
\begin{equation}
  \gamma(\Jf,f) = \frac{2}{f}\frac{\Re Z\on{f}}{\Rq} \on{\Gamma\on{\Jf-f} + \Gamma\on{-\Jf-f}}.
  \label{eq:PSD}
\end{equation}
Here the entire linear circuit in which the junction is embedded is described by an impedance $Z(f)$ in series with the junction. $\Gamma\on{\nu} \approx \frac{\Rq}{4} \Ic^2 P\on{h \nu}$ is the Cooper pair tunneling rate through the junction at bias $\nu$ with $\Rq=\frac{h}{4e^2}$ the superconducting resistance quantum and $P(E)$ the probability distribution for a tunneling Cooper pair to emit energy $E$ into the modes of $Z(f)$ \cite{Ingold1992}. It depends on $Z(f)$ and for small impedances ($\Re Z(f) \ll R_Q$) it can be approximated as $P(E\gg kT) \approx \frac{\pi}{E} \frac{\Re Z(E/h)}{\Rq}$ and $P(E\ll -kT) \approx 0$. Eq.~(\ref{eq:PSD}) is plotted in Fig.~\ref{fig:psd_gain}b for the sample design parameters and an effective Josephson energy $\Ic = \pq{20.2}{nA}$.



The amplification and absorption processes can be explained by calculating the effective admittance of the junction at frequency $f$. Probabilities for a photon impinging on the junction to be absorbed ($\gamma^-$) or being reflected while stimulating the emission of an additional photon ($\gamma^+$) are related to the spontaneous emission rate density Eq.~(\ref{eq:PSD}) \cite{Safi2011}:
\begin{equation}
  \gamma^\pm(\Jf,f) = \gamma(\Jf, \pm f).
\end{equation}
These rates result in an effective junction admittance $Y$ defined by
\begin{equation}
\Re Y(\Jf,f) = \frac{1}{4 \Re Z(f)} \left(\gamma^-(\Jf, f) - \gamma^+(\Jf, f) \right)\\
  \label{eq:junction-admittance}
\end{equation}
The real part of the junction admittance can become negative when the $\gamma^+$ term dominates. Then the reflection coefficient at the junction $G = \left|\frac{1-Z(f)Y(\Jf,f)}{1+Z(f)Y(\Jf,f)}\right|^2$ becomes $> 1$, corresponding to gain.

 Note however, that relation Eqs.~(\ref{eq:PSD}) and (\ref{eq:junction-admittance}), based on standard $P(E)$ theory, assume that tunneling events and photon generation are rare enough ($\gamma \ll 1$) that the photon modes in the circuit relax to equilibrium between tunneling events, a regime we have explored previously \cite{Hofheinz2011}. Here, on the contrary, we use a Cooper pair current large enough to drive the electromagnetic modes at $f_0$ and $f_1$ significantly out of equilibrium, as can be seen in Fig.~\ref{fig:psd_gain}a where we observe photon numbers well beyond 1. This is indeed necessary because in order to reach useful gain $G \gg 1$ one needs $Z(f)Y(\Jf,f) \rightarrow -1$, which implies violating the condition $\gamma \ll 1$. Therefore, Eqs.~(\ref{eq:PSD}) and (\ref{eq:junction-admittance}) only provide a qualitative description of amplification and noise. A quantitative description would require an extension of the $P(E)$ framework including non-thermal states of the electromagnetic environment.

 \textbf{Microwave calibration:} In order to calibrate PSD and VNA measurements, we place a Radiall R591763600 microwave switch (thermally anchored to the mixing chamber) between the bias-T and the chip (see Fig.~1). It connects the amplification chain either to the sample, to a short circuit or to $\pq{50}{\ohm}$ thermal loads, one thermally anchored to the mixing chamber, one to the still. The well-know thermal noise emitted by these resistors allows us to calibrate gain and noise of the amplification chain. In order to calibrate the attenuation of the input line we connect to the short circuit, reflecting the input signal and sending it to the already calibrated amplification chain.

 We have two possibilities to normalize VNA measurements. They can be normalized with respect to the reflection off the short circuit or with respect to microwave reflection off the sample at voltages where gain is close to 1 in Fig.~\ref{fig:psd_gain} and for fully frustrated SQUID. We observe very similar attenuation, indicating that loss between the microwave switch and the junction is negligible. However, the frequency dependence is not exactly the same due to parasitic reflections in the cable connecting the switch to the sample. We, therefore, use the latter calibration which cancels these modulations.

\textbf{Input noise:}
The output noise is measured as the difference between the noise the sample emits at bias $\Jf$ and at bias 0. It measures photon emission, i.e.\ it removes noise added by amplification chain and zero-point fluctuation. We therefore calculate the noise added by the sample as
\begin{equation}
  n_\mathrm{in} = \frac{n_\mathrm{out} + 1/2}{G} - \frac{1}{2} = \frac{n_\mathrm{out}}{G}  - \frac{1}{2}\left(1-\frac{1}{G}\right)
  \end{equation}
when it achieves a power gain $G$.

\textbf{Critical current:}
The nominal critical current is evaluated as $\pq{17.5}{nA}$ from prior measurements of the normal state resistance at $\pq{4}{K}$ and an estimated gap of $2\Delta=\pq{0.2}{meV}$ of our Aluminum junction using the Ambegaokar-Baratoff formula. The agreement with the critical current $\pq{20.2}{nA}$ obtained from fitting the gain signature is remarkably good given that we operate in a regime where Eq.~(\ref{eq:PSD}) is not strictly valid.

\textbf{Effective temperature:}
In order to estimate the effective temperature of the low-frequency electromagnetic environment we perform a PSD measurement similar to Fig.~\ref{fig:psd_gain} but at almost fully frustrated SQUID where Eq.~(\ref{eq:PSD}) is valid. We perform the integral
\begin{equation}
  \int_{\pq{4}{GHz}}^{\pq{8}{GHz}} \mathrm{d}f \gamma(\delta \nu + f, f) \propto P(h\delta \nu)
\end{equation}
for small $\delta \nu$, i.e.\ around the 1-photon process. It allows us to evaluate $P(h \delta \nu)$ around 0, which we fit with an effective temperature $T_{\mathrm{eff}} = \pq{54.7}{mK}$. We neglect here the second term in Eq.~(\ref{eq:PSD}): After integration it would lead to a smooth background which we find to be negligible, in agreement with $e^{-hf_0/kT_{\mathrm{eff}}} \approx 0.005 \ll 1$.


%

\textbf{Author contributions:} SJ, FB, AG performed the measurements and analyzed the data. MH, FP, DV designed and fabricated the sample, AG, MH, FB, RA, SJ, DH built the setup and wrote software. MH, SJ wrote the manuscript with input from all authors. 

\textbf{Acknowledgements:} We acknowledge fruitful discussions with J. Leppäkangas, C. Altimiras, M. Devoret, B. Kubala and J. Ankerhold as well as financial support from the Grenoble Nanosciences Foundation (grant WiQOJo), the European Union (ERC starting grant 278203 WiQOJo, ICT grant 218783 SCoPE) and the ANR (grants Masquelspec, AnPhoTEQ).

\end{document}